\documentclass[twocolumn,amsmath,amssymb,showpacs]{revtex4}
\begin{document}
\title{Exact scaling properties of a hierarchical network model}
\author{Jae Dong Noh}
\affiliation{Theoretische Physik, Universit\"at des Saarlandes, 66041,
Saarbr\"ucken, Germany}
\date{\today}

\begin{abstract}
We report on exact results for the degree $K$, the diameter $D$,
the clustering coefficient $C$, and the betweenness centrality $B$
of a hierarchical network model with a replication factor $M$.
Such quantities are calculated exactly with the help of recursion relations.
Using the results, we show that (i) the degree distribution follows
a power law $P_K \sim K^{-\gamma}$ with $\gamma = 1+\ln M /\ln
(M-1)$, (ii) the diameter grows logarithmically as $D \sim \ln N$ with
the number of nodes $N$, 
(iii) the clustering coefficient of each node is inversely proportional 
to its degree, $C \propto 1/K$, and the average clustering coefficient
is nonzero in the infinite $N$ limit, and (iv) the
betweenness centrality distribution follows a power law
$P_B \sim B^{-2}$. We discuss a classification scheme of scale-free networks
into the universality class with the clustering property
and the betweenness centrality distribution.
\end{abstract}
\pacs{89.75.-k, 89.75.Da, 05.10.-a}
\maketitle

A network structure of complex systems has been attracting much research
interest~\cite{AB02}.
Since the works in Refs.~\cite{Watts98,AHB99}, 
it has been recognized that complex systems
have neither a regular nor a random network structure.  
Instead, complex networks found in various areas have a scale-free~(SF)
structure characterized with a power-law 
distribution of the degree,
see Ref.~\cite{AB02} and references therein. A possible mechanism for
the emerging SF structure is suggested by the Barab\'asi-Albert~(BA)
model~\cite{BA99}. However, the BA network lacks the clustering property
which is observed in many real networks.
To fill this gap, a hierarchical~(H) network model was introduced as a model
for SF networks with a clustering property~\cite{Ravasz02a}. 
It is observed that the H network displays a scaling law
$C\sim K^{-1}$ between the clustering coefficient $C$ of a node and its
degree $K$ and that the average value of $C$ does not vanish in the infinite
size limit. The clustering property is also observed in some metabolic 
networks~\cite{Jeong00}, which is regarded as an evidence for a hierarchical
structure~\cite{Ravasz02a,Ravasz02b}.

On the other hand, the clustering property does not necessarily 
imply a hierarchical structure. 
Some SF network models~\cite{Holme,Klemm,Szabo,Dorogovtsev01} 
have the clustering property, but it is not apparent whether they
have a hierarchical structure. 
Therefore, it is desirable to study other scaling properties 
of the H network model to establish the universality class for the 
hierarchical network.
This is the purpose of the current work. We derive analytically
exact scaling laws for the degree distribution, the diameter $D$, and the
clustering coefficient~(abbreviated to CC), and the betweenness 
centrality~(abbreviated to BC) $B$~\cite{Ravasz02a}. 
The CC is a measure of local connectivity or modularity, while
the BC reflects a global property~\cite{Newman,Goh}. 
The scaling property of both quantities
characterizes the universality class for the H network.

The H network is characterized by $G$ (the number of generations) and
$M$~(a replication factor). The network of the $G$th generation,
denoted as ${\cal N}_G$, has $N=M^G$ nodes, which will be labelled by 
the coordinate $G$-tuple of integers $[\mbox{\boldmath $x$}_G]\equiv 
[x_G\ldots x_1]$ with $0\leq x_i <M$. 
The network is defined recursively. 
The first generation consists of one central node $[0]$, which will be 
referred to as a hub, and $(M-1)$ peripheral nodes 
$[y]$ with $1\leq y <M$. 
All nodes are fully connected with each other. 
Suppose one has ${\cal N}_{G-1}$, each node of which
is assigned to a coordinate $[\mbox{\boldmath $x$}_{G-1}]$.
In the next generation, $(M-1)$ copies are 
added to the network and all their peripheral nodes are connected to 
the hub of the original unit. 
Nodes in the original unit are assigned to $[0\mbox{\boldmath $x$}_{G-1}]$,
and the copies to $[y\mbox{\boldmath $x$}_{G-1}]$
with $y=1,\ldots,M-1$, respectively. It leads to the network ${\cal N}_G$. 
Nodes $[y_{G}\cdots y_1]$ with $y_i\neq 0$ for all $i$ are the 
peripheral nodes and $[{\bf 0}_G]$ is the hub of ${\cal N}_G$.

With the coordinate system, geometrical
properties of the network can be studied combinatorially.
In general, the node connectivity follows the rule:
\begin{itemize}
\item[\bf C1:]{ $[x_G \cdots x_{l+1} y_l \cdots y_1 ]  \leftrightarrow
[x_G \cdots x_{l+1} y_{l}\cdots y_{k+1} \mbox{\boldmath $0$}_k]$ \\
$(1\leq l\leq G \mbox{ and } 1\leq k\leq l)$}
\item[\bf C2:]{
$[x_G\cdots x_{2}\ y_1 ] \leftrightarrow [x_G\cdots x_{2}\ y'_1]$
for  $y_1\neq y'_1 $}
\end{itemize}
Above and hereafter,
we use $x$ for a dummy variable ranging from 0 to
$(M-1)$ and $y$ for one ranging from 1 to $(M-1)$. 
The rule {\bf C1} comes from the fact that all peripheral nodes of 
${\cal N}_k$ are connected to
the hub $\mbox{\boldmath $0$}_k$ during the replication, and {\bf C2}
from the fact that all peripheral nodes of ${\cal N}_1$ are
fully inter-connected.

For a convenience, we classify nodes into four sets:
(a) ${\mathbb P}$ for peripheral nodes $[y_G \cdots y_1]$,
(b) ${\mathbb{LP}l}$~(stands for Locally Peripheral) ($1\leq l <G$) 
for nodes of a form $[x_G\cdots x_{l+2}\ 0\ y_l\cdots y_1]$,
(c) $\mathbb{LH}l$~(Local Hub) ($1\leq l<G$) for nodes of a form
$[x_G\cdots x_{l+2} y_{l+1}\ \mbox{\boldmath $0$}_l]$, and
(d) $\mathbb{H}$ for the hub $[\mbox{\boldmath $0$}_G]$.
The size $S$ of each set is given in Table~\ref{table1}.
\begin{table*}
\caption{The degree $K$, the clustering coefficient $C$, the
partial betweenness centrality $B^0$, and the betweenness centrality $B$
of a node in each set and the number of nodes $S$ in each set.}
\label{table1}
\begin{ruledtabular}
\begin{tabular}{cccccc}
set    &  $S$  &  $K$ & $C$ & $B^0$ & $B$ \\ \hline
$\mathbb{P}$           &  $(M-1)^G$ & $G+(M-2)$ & $
\frac{(M-2)(2G+M-3)}{(G+M-2)(G+M-3)}$ &
$\left(\frac{M}{M-1}\right)^{G-1}-1$ & $ \simeq 2 M^G \left(
\frac{M(M^2-1)}{M^3+1}\right) \left\{
\left(\frac{M}{M-1}\right)^{G-1}-1\right\}$ \\ [3mm]
$\mathbb{LP}l$      &  $ (M-1)^l M^{G-(l+1)}$ & $l+(M-2)$ &
$\frac{ (M-2)(2 l + M-3)}{ (l+M-2)(l+M-3)}$ &
$\left(\frac{M}{M-1}\right)^{l-1}-1$ &
$ \simeq 2 M^G \left\{ \left(\frac{M}{M-1}\right)^{l-1}-1\right\}$,
for $l\ll G$ \\ [3mm]
$\mathbb{LH}l$      &  $ (M-1) M^{G-(l+1)}$  & $\sum_{n=1}^l (M-1)^n$ &
$\frac{(M-2)}{\sum_{n=1}^l (M-1)^n - 1}$ &
$M^{l-1}-1$ &  $\simeq 2M^G (M^{l-1}-1)$, for $l\ll G$\\ [3mm]
$\mathbb{H}$           &  $ 1 $                  & $\sum_{n=1}^G (M-1)^n$ &
$\frac{(M-2)}{\sum_{n=1}^G (M-1)^n - 1}$ & $0$&
$ \simeq \left\{ \frac{2}{M+1} + \frac{M^2(M-2)}{M^2-1} \ln
\frac{M^2}{M^2-M+1}\right\} M^{2G}$ \\ [3mm]
\end{tabular}
\end{ruledtabular}
\end{table*}

{\em Degree distribution --- } Using the connection rules, one
can easily enumerate the degree, which is the number of neighbors,
of each node. All nodes in the same set have the same degree, which are
presented in Table~\ref{table1}.
The nodes in $\mathbb{P}$ and $\mathbb{H}$ contribute to the degree 
distribution $P_K$ at isolated points of $K$.
For nodes in $\mathbb{LP}l$,
it is given by $P_K = M^{-G} S_l | \Delta l / \Delta K_l |$ with $\Delta l =
1$ and $\Delta K_l \equiv K_{l+1}-K_l = 1$. Using $K$ and $S$ in
Table~\ref{table1}, we find that $P_K \sim \exp[K\ln(1-1/M)]$.
On the other hand, nodes in $\mathbb{LH}l$ has $K_l \sim (M-1)^l$ and $S_l\sim M^{-l}$.
Hence, the degree distribution follows a power law 
$P_K \sim K^{-\gamma}$ for $M>2$  with $\gamma = 1 + \frac{\ln M}{\ln (M-1)}$.
The nodes in $\mathbb{LH}l$ have larger degrees than those in $\mathbb{LP}l$. Therefore,
the total degree distribution follows the power law with
the exponent $\gamma$ in the tail region.
In particular, the hub has the largest degree, which scales
as $K_{hub} \sim N^{1/(\gamma-1)}$.

{\em Shortest path and diameter ---}
We first consider a shortest path from an arbitrary node
$[x_G\cdots x_1]$ to the hub $[\mbox{\boldmath $0$}_G]$.
One can reach the hub by flipping the coordinate using the rules
{\bf C1} and {\bf C2} successively.
The shortest path has the minimum number of steps, which is called the
distance. Note that a step using {\bf C2} always leads to a detour.
Hence it must not be used in finding a shortest path to the hub.
Following {\bf C1}, one may flip the {\em consecutive} 
lowest~(including $x_1$) figures that are all zero
to nonzero values, or vice versa. The whole consecutive lowest figures
$x_i \cdots x_1$ with $x_{1\leq j\leq i}=0~(x_{i\leq j\leq i}\neq 0)$ 
and $x_{i+1}\neq 0$~($x_{i+1}=0$) will
be referred to as a zero~(nonzero) {\em domain}.
Then, one can reach the hub in minimal steps by flipping a zero/nonzero
domain to a nonzero/zero domain successively.
The domain size increases at each flip until one reaches the
hub, a zero domain of size $G$.  Since a zero domain can be flipped to
any nonzero domain, the shortest path has large degeneracy in general.

The process resembles domain-coarsening in magnetic systems.
To complete this analogy, we map
the coordinate $[x_G\cdots x_1]$ onto a spin state of the $(Q=M)$-state
Potts model in one-dimensional lattice of size $G$ and assign the energy
with the Hamiltonian
\begin{equation}
{\cal H} = \sum_{i=1}^G \left\{ 1 - \delta( \delta(x_i,0), \delta(x_{i+1},0) )
\right\} ,
\end{equation}
where $\delta(,)$ is the Kronecker delta symbol and $x_{G+1}\equiv0$ is a fixed
ghost spin.
Then the distance between a node to the hub is given by the energy 
of the spin state.
Therefore the mean node-to-hub distance
is given by the average energy of the spin
system in the infinite temperature limit:
$D_H(G) = -\left[ {\partial \ln Z(\beta;G)}/{\partial \beta}
\right]_{\beta=0}$
with the partition function $Z(\beta;G) = \sum_{\mbox{\boldmath $x$}_G}
e^{-\beta {\cal H}[\mbox{\boldmath $x$}_G]}$.
It can be calculated using a transfer matrix
method. After some algebra, we obtain that
\begin{equation}
D_H(G) = \frac{2(M-1)}{M^2}G + \frac{(M-1)(M-2)}{M^2} .
\end{equation}

A shortest path between arbitrary nodes $[\mbox{\boldmath $x$}_G] =
[x\mbox{\boldmath $x$}_{G-1}]$ and $[\mbox{\boldmath $x'$}_G] =
[x'\mbox{\boldmath $x'$}_{G-1}]$ can be found recursively. If $x\neq
x'$, all paths connecting them pass through the hub
$[\mbox{\boldmath $0$}_G]$. So, the shortest path is given
by a shortest path from one node to $[\mbox{\boldmath $0$}_G]$ followed by
a shortest path from $[\mbox{\boldmath $0$}_G]$ to the other. If
$x=x'$, one can restrict the shortest path within a sub-network
of all nodes $[x''\mbox{\boldmath $x''$}_{G-1}]$ with $x''=x$,
for a path utilizing other nodes with $x''\neq x$
does not reduce a path length~\cite{comment1}.
Note that the sub-network with all links to other nodes in ${\cal N}_{G}$
disabled has that same structure as ${\cal N}_{G-1}$.
Therefore, the distance satisfies the recursion relation
\begin{eqnarray}
d([\mbox{\boldmath $x$}_G],[\mbox{\boldmath $x'$}_G])\! &=&\!
\delta'(x,x') \left\{ d([\mbox{\boldmath $x$}_G],
[\mbox{\boldmath $0$}_G]) +
d([\mbox{\boldmath $x'$}_G],[\mbox{\boldmath $0$}_G]) \right\}
\nonumber \\
&+& \delta(x,x')\ d ( [\mbox{\boldmath $x$}_{G-1}],[\mbox{\boldmath
$x'$}_{G-1}])
\end{eqnarray}
where $\delta'(,) \equiv 1-\delta(,)$ for short. Summing up over all
node pairs, we obtain the recursion relation
$D(G) = 2M^{-1}(M-1) D_H(G) + M^{-1} D(G-1)$
for the diameter~(mean node-to-node distance)
with the solution
\begin{equation}
D(G) = \frac{4(M-1)G}{M^2} + \frac{2(M-3)}{M} -
\frac{( M^2-M-4)}{M^{G+1}} .
\end{equation}
In the infinite $N=M^G$ limit, we find that $D\simeq 2 D_H$ and that
the diameter scales logarithmically with $N$. It 
is a characteristics of the hierarchical network.
For conventional~(non-hierarchical) SF networks, the diameter scales
sub-logarithmically for $\gamma \leq 3$~\cite{Cohen02,Noh02}.

{\em Clustering coefficients ---}
The CC of a node with $K$ neighbors is given by 
$C = 2 N_e / K (K-1)$, where $N_e$ is the number of existing edges between
$K$ neighbors. Using the connection rules, it is straightforward to
calculate the CC of each node. Nodes in the same set
have the same value of $C$, which are presented in Table~\ref{table1}.

Using the results in Table~\ref{table1}, the degree dependence of the CC 
is easily obtained.
For nodes in $\mathbb{P}$ and $\mathbb{LP}l$, 
we obtain that $C(K) = (M-2)(2K-M+1)/K(K-1)$ with $M-1\leq K \leq G+(M-2)$.
So, for large $K\gg M$, their CC's
are inversely proportional to the degrees, $C(K) \simeq 2(M-2) / K$.
The CC's of nodes in $\mathbb{LH}l$ and the hub are exactly
given by $C(K) = (M-2)/(K-1)$. The scaling law $C\simeq c K^{-1}$ holds for 
both cases, but the coefficient $c$ differs by a factor of two~(cf.
Fig.~2(b) in Ref.~\cite{Ravasz02a}). 

The average CC is given by $\bar{C}\equiv M^{-G}
( S_\mathbb{P} C_\mathbb{P} + \sum_l (S_{\mathbb{LP}l} C_{\mathbb{LP}l} + S_{\mathbb{LH}l} C_{\mathbb{LH}l}) + S_\mathbb{H} C_\mathbb{H})$. In the
infinite size limit~($G \rightarrow \infty$), it converges to a {\em nonzero} 
value
\begin{eqnarray}
\bar{C} &=& (1-\frac{2}{M}) \sum_{l=1}^\infty
\frac{(2l + M-3)(1-1/M)^l }{(l+M-2)(l+M-3)} \nonumber \\
&+&
(1-\frac{1}{M}) \sum_{l=1}^\infty \frac{(M-2)^2}{M^{l} ( (M-1)^{l+1} - 2M+3)}.
\end{eqnarray}
Numerically $\bar{C} = 0.719282...$ and $0.741840...$ for $M=4$ and $5$,
respectively. $\bar{C}$ converges to 1 as $M\rightarrow \infty$.

{\em Betweenness centrality --- }
The BC of a node
is the sum of weights of shortest paths between all node pairs
that pass through the node.
For a given node pair, all degenerate shortest paths connecting them are
weighted with the inverse of the degeneracy.
First of all, we calculate a so-called partial BC
$B^0$, which is obtained from a partial sum over
all shortest paths between the hub and the others.
It is calculated easily, since all shortest paths
can be constructed using the domain-coarsening picture.

Again, each node in the same set has the same value of $B^0$.
(a) A node $u = [y_G\cdots y_1] $ in $\mathbb{P}$ may belong to a shortest 
path to the hub from nodes $[y_G \cdots y_{l+2} 0 x'_l \cdots
x'_1]$ with $0\leq l\leq G-2$ and arbitrary $x'_i$.
The domain-coarsening process leads them to $[y_G\cdots y_{l+2}
{\bf 0}_{l+1}]$ at an intermediate step. Then, the zero domain ${\bf
0}_{l+1}$ flips to a nonzero domain in the next step
with probability $(M-1)^{-(l+1)}$ passing through the node $u$.
Hence, each node contributes $(M-1)^{-(l+1)}$ to $B^0_u$.
Summing up all contributions, we obtain the result in Table~\ref{table1}.
(b) A node $[x_G\cdots x_{l+2} 0
y_l\cdots y_1]$ in $\mathbb{LP}l$ may belong to a shortest path
from nodes $[x_G\cdots x_{l+2} 0 y_l\cdots y_{m+2} 0 x'_m\cdots x'_1]$
with $0\leq m \leq l-2$ and arbitrary $x'_i$ to the hub. Following the
same idea as in (a), one can easily obtain $B^0$, see
Table~\ref{table1}. (c) A node $u = [x_G\cdots x_{l+2} y_{l+1} \mbox{
\boldmath $0$}_l]$ in $\mathbb{LH}l$ belongs to ``all" shortest paths from
nodes $[x_G\cdots x_{l+2}y_{l+1} 0 x'_{l-1}\cdots x'_{1}]$
with arbitrary $x'_i$ except for $u$ itself. So, $B^0_u =
M^{l-1}-1$. (d) Trivially, $B^0=0$ for the hub.

By eliminating the parameter $l$ in Table~\ref{table1}, we obtain that
$B^0 (K) = (M/(M-1))^{K-M+1}$ for nodes in set $\mathbb{LP}l$.
It is diverging exponentially with $K$. On the other hand, $P_K$ for nodes 
in $\mathbb{LP}l$ decays exponentially. So,
$P_{B^0}$ decays algebraically. Explicitly it is obtained from
$P_{B^0} = M^{-G} S_l | \Delta l / \Delta B_l^0|$, which yields that
\begin{equation}\label{P_B0p}
P_{B^0} = \left(\frac{M-1}{M}\right)^2 \frac{1}{(B^0+1)^2} \ .
\end{equation}
We obtain the relation
$B^0(K) = -1+M^{-1} (1+K (M-2)(M-1)^{-1})^{\gamma-1}$ 
for nodes in the set $\mathbb{LH}l$. 
So the distribution is given by
\begin{equation}\label{P_B0h}
P_{B^0} = \frac{1}{M^2} \frac{1}{(B^0+1)^2} \ .
\end{equation}
Therefore, the partial betweenness centrality has the power-law
distribution with the exponent $2$.

The mean node-to-node distance was 
obtained using the mean node-to-hub distance. We apply a similar idea for
the BC. Introduce a character function $\chi_G([\mbox{
\boldmath $x'$}_G],[\mbox{\boldmath $x''$}_G];[\mbox{\boldmath $x$}_G])$
to denote the fraction of paths passing through $[\mbox{\boldmath $x$}_G]$
among all shortest paths between $[\mbox{ \boldmath $x'$}_G]$ and
$[\mbox{ \boldmath $x''$}_G]$ in ${\cal N}_G$.
The $[\mbox{\boldmath $x$}_G]$ dependence will be assumed implicitly.
Then, the BC of a node $[\mbox{\boldmath $x$}_G]=
[x\mbox{\boldmath $x$}_{G-1}]$ can be written formally as
$B_{[\mbox{\boldmath $x$}_G]} = \sum_{ \mbox{\boldmath $x$}'_G,
\mbox{\boldmath $x$}''_G} \chi_G ( [\mbox{
\boldmath $x$}'_G],[\mbox{\boldmath $x$}'_G])$.
The hub will be considered separately later, and we assume that
$[\mbox{\boldmath $x$}_G]$ is not the hub for the time being.
Decompose the sum $\sum_{\mbox{\boldmath $x'$}_G}$ into
$\sum_{x'} \sum_{\mbox{\boldmath $x$}'_{G-1}}$ and similarly for
$\mbox{\boldmath $x$}''_G$. When $x'\neq x''$, all shortest paths between
$[x'\mbox{ \boldmath $x$}'_{G-1}]$ and
$[x''\mbox{ \boldmath $x$}''_{G-1}]$ pass through the hub,
which yields that
$\chi_G([x'\mbox{ \boldmath $x$}'_{G-1}],[x''\mbox{ \boldmath $x$}''_{G-1}])
= \chi_G([x'\mbox{ \boldmath $x$}'_{G-1}],[{\bf 0}_G]) +
\chi_G([x''\mbox{ \boldmath $x$}''_{G-1}],[{\bf 0}_G])$.
When $x'=x''$ and $[\mbox{\boldmath $x$}_G]$ is not the hub, the summands
are nonzero only when $x'=x''=x$.
Using these properties, we obtain that
\begin{eqnarray}\label{BC}
B_{[\mbox{\boldmath $x$}_G]} &=& 2(M-1)M^{G-1} B^0_{[\mbox{\boldmath $x$}_G]}
\nonumber \\
&+& \sum_{\mbox{ \boldmath $x$}'_{G-1},\mbox{ \boldmath $x$}''_{G-1}}
\chi_G( [x\mbox{\boldmath $x$}'_{G-1}],[x\mbox{\boldmath $x$}''_{G-1}]) .
\end{eqnarray}

One might be tempted to identify the second term as $B_{[\mbox{\boldmath
$x$}_{G-1}]}$, i.e., the BC defined on ${\cal N}_{G-1}$.
However this is not correct for $x\neq 0$, since some pairs of
$[\mbox{\boldmath $x$}'_{G-1}]$ and $[\mbox{\boldmath $x$}''_{G-1}]$ may have
degenerate shortest paths passing through the hub $[{\bf 0}_G]$ which is not
present at ${\cal N}_{G-1}$. This was explained when we discussed the
diameter.
For example, $[y1111]$ and $[y1222]$ with nonzero $y$
are connected via $[y1000]$ and $[y0000]$, and also via $[00000]$,
so each path has the weight $1/3$. However, if one ignores the path
via $[00000]$, the other paths would have weight $1/2$, and hence the nodes
$[y1000]$ and $[y0000]$ would have larger value of the BC.

Therefore, the second term in Eq.~(\ref{BC}) should be written as a sum of
$B_{[\mbox{\boldmath $x$}_{G-1}]}$ and a quantity that compensate
for the change in the degeneracy of the shortest paths.
As the example showed, the compensation is necessary only for nodes 
of the form 
$[\mbox{\boldmath $x$}_G]= [y_G\cdots y_{l+1}\mbox{\boldmath $0$}_l]$
with $1\leq l < G$. 
For such nodes, after enumerating all degeneracy carefully, 
Eq.~(\ref{BC}) becomes
\begin{eqnarray}\label{recursion1}
B_{[\mbox{\boldmath $x$}_G]} &=& 2(M-1) M^{G-1} B^0_{[\mbox{\boldmath $x$}_G]}
+ B_{[\mbox{\boldmath $x$}_{G-1}]} \nonumber \\
&-& \sum_{k=1}^{l-1} \frac{(M-2) (M-1)^k M^{2(l-k)}}{ (G-l+k)(G-l+k-1) } \ .
\end{eqnarray}
The hub gains from such degenerate shortest paths, 
which
lead to a similar recursion relation
\begin{eqnarray}\label{recursion2}
B_{[\mbox{\boldmath $0$}_G]} &=& (M-1)( M^{2G-1} - 2 M^{G-1})
+ B_{[\mbox{\boldmath $0$}_{G-1}]} \nonumber \\
&+& (M-2) M^{2G} \sum_{k=2}^{G-1} \frac{(M-1)^k }{k M^{2k}} \ .
\end{eqnarray}
For other nodes, we have the simple relation
\begin{equation}\label{recursion3}
B_{[\mbox{\boldmath $x$}_G]} = 2(M-1) M^{G-1} B^0_{[\mbox{\boldmath
$x$}_G]} + B_{[\mbox{\boldmath $x$}_{G-1}]}\ .
\end{equation}

With the recursion relations, we now readily calculate the 
BC of all nodes exactly. Since the exact expressions are lengthy, we
present only the leading order contribution in the large $G$ limit in
Table~\ref{table1}.
Note that $B \simeq 2 M^G B^0$ for most nodes  
in $\mathbb{LP}l$ and $\mathbb{LH}l$ in the large $G$ limit.
The proportionality relation implies that the
BC follows the same power-law distribution as $B^0$:
\begin{equation}
P_B \sim B^{-2}  \ .
\end{equation}

{\em Summary and discussions ---}
We have studied the exact scaling properties of the H network introduced in
Ref.~\cite{Ravasz02a}.
We have shown that the H network has the clustering property:
The average value of the CC is nonzero in the infinite network size limit and 
the CC exhibits the scaling law $C\sim K^{-\beta}$
with $\beta=1$. We have also shown that the BC
follows the power-law distribution $P_B \sim B^{-\eta}$
with $\eta=2$. Both scaling properties characterize 
the H network model.

The BC proved to be useful in classifying SF networks into 
the universality class. The BC distribution exponent
is universal and has the value either $\eta\simeq 2.2$ in
the class I or $\eta\simeq 2.0$ in the class II~\cite{Goh}. 
Combining the scaling properties
of the CC and the BC, we suggest that there exist four classes, 
that is, I-C, I-NC, II-C, and II-NC~(``C" for clustered and ``NC" for 
non-clustered networks)~\cite{comment2}.
The H network model then belongs to the 
class II-C. The Internet at the autonomous system level 
and some metabolic networks of archaea display both scaling behaviors with 
$\beta\simeq 1.0$~\cite{Ravasz02a} and $\eta\simeq 2.0$~\cite{Goh}. 
So they belong to the same class II-C, which is a stronger 
evidence for a hierarchical structure~\cite{Ravasz02a}.
The BA network with $m=1$~\cite{BA99} and the deterministic tree 
network~\cite{Jung} have $C=0$ and $\eta=2$~\cite{Oh}, 
thus they are members of the class II-NC.
The BA network with $m\ge 2$ has vanishing CC~\cite{Klemm,Szabo} 
and $\eta\simeq 2.2$~\cite{Goh}, and belongs to the class I-NC.

Literatures suggest that the metabolic networks of bacteria and 
eukaryotes~\cite{Goh,Ravasz02b} and the coauthorship network in the field 
of neuroscience~\cite{Goh,Goh02} might belong to the class I-C. With the
existence of the class I-C, the clustering property does not necessarily 
imply the hierarchical structure. Therefore, it is important to establish
the class I-C firmly. Further studies on the BC distribution
in model networks with the clustering property, such as the 
Holme-Kim model~\cite{Holme} and
Klemm-Egu\'iluz model~\cite{Klemm}, are required. 
Further studies are also necessary
to reveal the similarity/dissimilarity between the metabolic networks of 
archaea and those of bacteria and eukaryotes, which have different BC
distributions.

Goh {\it et al.}~\cite{Goh} suggested that the topology of shortest pathways 
be a universal characteristics of SF networks. 
They found a chain-like structure for networks in the class II~($\eta=2$).
It was presumed that the chain-like structure leads to a
linear mass-distance relation $m(d) \simeq  A d$
where $m(d)$ is the mean number of nodes along all shortest paths
between a node pair separated by a distance $d$. 
We also found that the shortest pathways of the H network have a chain-like 
structure. 
For example, a set of all shortest paths from $[001010]$ to the 
hub in ${\cal N}_6$ 
is given by $[001010] \stackrel{1}{\rightarrow} [00101y] 
\stackrel{2}{\rightarrow} [001{\bf 0}_3] \stackrel{3}{\rightarrow} 
[001yyy] \stackrel{4}{\rightarrow} [{\bf 0}_6]$ with arbitrary nonzero 
$y$'s. It has a chain-like structure, but the steps 1 and 3 
introduce blobs whose size increases {\em exponentially} as one proceeds.
We could show that $m(d)$, averaged over all nodes separated by the distance
$d$ from the hub, satisfies an inequality $m(d) \ge a (M-1)^d$ 
with a positive constant 
$a$~\cite{noh_unpub}. So a topological characterization other than the
mass-distance relation is necessary to characterize the chain-like structure
observed in the class II-C.

This work has been financially supported by the
Deutsche Forschungsgemeinschaft (DFG).
The author thanks H. Rieger and B. Kahng for useful discussions.


\begin{references}
\bibitem{comment1} If $x=x'\neq 0$, there may exist degenerate shortest
paths passing through the hub which does not belong to the sub-network.
As far as the distance is concerned, such degenerate paths can be
ignored.
\bibitem{comment2} A clustered network may have 
the exponent $\beta\ne 1$~\cite{Ravasz02a}.
\end{references}
\end{document}